\documentclass[a4paper]{article}      

\usepackage{amsmath,amssymb,amsfonts,amsthm} 
\usepackage{graphicx}

\usepackage{verbatim}

\usepackage[margin=1cm,font=small]{caption}

\usepackage{layout} 

\newcommand{\boldomega}{{\boldsymbol{\omega}}}

\newcommand{\ProbaStar}{{\mathsf{P}_{\hspace{-0.05cm}\star}}}
\newcommand{\MeanStar}{{\mathsf{E}_{\star}}}

\newtheorem{Theorem}{Theorem}

\newtheorem{Lemma}{Lemma}

\newcommand{\Proof}{\noindent\emph{Proof.\hspace{0.15cm}}}

\newcommand{\Remark}{\noindent\textbf{Remark.\hspace{0.15cm}}}
\newcommand{\Remarks}{\noindent\textbf{Remarks.\hspace{0.15cm}}}

\newcounter{CompteurRemark}

\newcommand{\Lp}{\mathrm{L}}

\newcommand{\R}{\mathbb{R}}

\newcommand{\Z}{\mathbb{Z}}
\newcommand{\N}{\mathbb{N}}

\newcommand{\dd}{\mathrm{d}}

\newcommand{\ed}{\mathrm{e}}

\newcommand{\Mean}{\mathsf{E}}
\newcommand{\Proba}{\mathsf{P}}

\newcommand{\multiplication}{\, . \,}

\begin{document}

\title{
Drastic fall-off of the thermal conductivity for disordered lattices in the limit of weak anharmonic interactions
}
\author{
Fran\c cois Huveneers\footnote{
CEREMADE, 
Universit\' e de Paris-Dauphine,
Place du Mar\' echal De Lattre De Tassigny,
75775 PARIS CEDEX 16 - FRANCE.
E-mail: huveneers@ceremade.dauphine.fr.
Supported by the European Advanced Grant Macroscopic Laws and Dynamical Systems (MALADY) (ERC AdG 246953)}
}
\date{}

\maketitle 

\begin{abstract}
\noindent
We study the thermal conductivity, at fixed positive temperature, of a disordered lattice of harmonic oscillators,
weakly coupled to each other through anharmonic potentials. 
The interaction is controlled by a small parameter $\epsilon >0$. 
We rigorously show, in two slightly different setups, that the conductivity has a non-perturbative origin.
This means that it decays to zero faster than any polynomial in $\epsilon$ as $\epsilon\rightarrow 0$.
It is then argued that this result extends
to a disordered chain studied by Dhar and Lebowitz \cite{dha}, and to a classical spins chain recently investigated by Oganesyan, Pal and Huse \cite{oga}.
\end{abstract}

\vspace{11cm}

\pagebreak

\section{Introduction}\label{sec: Introduction}

It is generally admitted that the thermal properties of solids can be derived from molecular dynamics, 
but it remains to this day a widely open conjecture, at least from a mathematical point of view \cite{bon}\cite{dol}\cite{liv}.
Indeed, the few hamiltonian systems that can be handled analytically to a large extent, 
also appear to have a very pathological thermal behavior.
So is it for the ordered harmonic chain \cite{rie}, for the Toda lattice (see Section 6.3 in \cite{ber3}), and for a one-dimensional system of colliding particles \cite{rya}. 
As we will see soon, disordered one-dimensional harmonic chains also fall into this category. 

This said, much progress have been recently accomplished in the case of a harmonic crystal that is weakly perturbed by means of anharmonic interactions \cite{lef}\cite{aok}\cite{luk}.
In particular, a Boltzmann-like equation for phonons can be, at least formally, derived in the limit where the interaction vanishes,
after that time has been properly rescaled. 
A similar strategy is implemented in \cite{dol}\cite{liv} to another kind of system. 
In these examples, the conductivity can thus be understood in a perturbative regime. 

Besides anharmonicity, impurities constitute another possible mechanism destroying ballistic transport of energy. 
In disordered harmonic crystals, one assumes that the mass of each atom is random \cite{rub}\cite{cas}.
The dynamics is integrable and, in dimension one, the conductivity of disordered harmonic chains is by now rather well understood \cite{ver}\cite{aja}.
In particular, in presence of everywhere onsite pinning, Anderson localization of the eigenmodes forbids any transfer of energy \cite{ber2}.  
Though still integrable in the mathematical sense, the dynamics is much less understood in higher dimensions. 
Normal conductivity is expected in dimension three if the disorder is small enough \cite{cha}. 

We here look at the effect of both disorder and anharmonic interactions on a lattice of pinned harmonic oscillators. 
Three studies have attracted our attention.  
First, Dhar and Lebowitz \cite{dha} conclude from numerical experiments that the conductivity of a one-dimensional pinned disordered harmonic chain becomes positive as soon as
some anharmonicity is added, destroying thus localization. 
The same conclusion is then reached by Oganesyan, Pal and Huse \cite{oga}, which study a chain of classical spins. 
Moreover, they observe a very rapid fall-off of the conductivity as the parameter $\epsilon > 0$ (called $J$ in \cite{oga}) controlling the interaction goes to zero. 
Finally, Basko \cite{bas} argues that the conductivity in a comparable system can be understood as an effect of Arnold diffusion. 
He also gives an expression for the conductivity, 
predicting a decay with $\epsilon\rightarrow 0$ that is faster than any power law
(equations (3.6) in \cite{bas} with $\rho$ in place of $\epsilon$). 

Let us describe our contribution. 
We study a $d$-dimensional lattice of $1$-dimensional harmonic oscillators with random eigenfrequencies, as described in detail in Section \ref{sec: Model}. 
For $\epsilon = 0$, they just evolve independently from each other, 
but, for $\epsilon >0$, each oscillator is coupled to its neighbors through an anharmonic potential of strength $\epsilon$.  

Since the rigorous analysis of the conductivity of this system is likely out of reach, we study it in two slightly different simplified setups.
Let $n >> 1$ be some large integer that we fix.
The first simplifying procedure consists in 
following the dynamics only up to very large time-scales of order $\epsilon^{-n}$, and study the conductivity in the strict weak coupling limit $\epsilon \rightarrow 0$,
as in \cite{aok}\cite{luk}\cite{dol}\cite{liv} for $n=2$.
Alternatively, we can perturb the hamiltonian dynamics by  a stochastic noise that preserves the energy of individual atoms, 
and possibly models some chaotic behavior of the system.
This noise is however chosen such that it only becomes sensitive at time-scales of order $\epsilon^{-n}$. 
It destroys hypothetical ballistic motion of energy only after such long times. 
These two approaches are very close in spirit.  

Our results are stated in Section \ref{sec: Conductivity}. 
Let $0 < \beta < + \infty$ be some given inverse temperature. 
Let $\kappa (\epsilon)$ be the Green-Kubo conductivity at inverse temperature $\beta$ (see \eqref{defintion Green-Kubo} below).
Theorem \ref{the: faster than any power law} deals with the first setup. 
Let $m$ be an integer such that $1 << m \le n$.  
It asserts that that $\epsilon^{-m}\kappa (\epsilon)$ vanishes in the weak coupling limit, 
no matter how large the integers $m \le n$ are taken.
We believe this to be a strong indication that $\kappa (\epsilon) = \mathcal O (\epsilon^m)$ for every $m\ge 1$.
Theorem \ref{the: faster than any power law bis} deals with the second setup, 
and asserts that, in that case, $\kappa (\epsilon) = \mathcal O (\epsilon^{n+2})$.
It extends, in a simpler model, results obtained in \cite{ber}\cite{ber2}, where the noise was dominating on the anharmonic interactions.
Incidentally, a comparison of our work with \cite{ber} and \cite{ber2} 
reveals that the effect of noise and that of anharmonic potentials can sometimes be very different in some respects.

Since the conductivity is defined at positive temperature, 
we find it hard to establish a clear connection between our work and the numerous existing results dealing with the possible dispersion of a finite energy packet at infinite volume
(among many other references, let us mention \cite{fro}\cite{pos}\cite{mac} for mathematical results, or \cite{mul} for numerical ones).
Still, in the perturbative regime that we consider, our results actually follow from a purely local analysis ; 
they are based on the lack of resonances in the system, a common phenomenon well described for example in Section 2 of \cite{hai}.


Theorems \ref{the: faster than any power law} and \ref{the: faster than any power law bis} are proven in Section \ref{sec: Proof of Theorem}.
In Section \ref{sec: Related models}, we show, without providing a full mathematical treatment, how to adapt our line of reasoning to obtain similar conclusions for
the disordered chain of \cite{dha} and the spin chain of \cite{oga}, 
in agreement with the observations of \cite{oga} and the predictions of \cite{bas}.
While the comparison with our model is completely straightforward in the case of the spin chain, if some assumption on the distribution of magnetic field is added,
at least some technical work would be required to get a true mathematical proof for the chain of \cite{dha}. 
We finally suggest a connection with a chain of weakly coupled identical strongly anharmonic oscillators.

\section{Model}\label{sec: Model}

Let $N \ge 3$, and let $\Z_N$ be the set of integers modulo $N$.
Let also $d \ge 1$.
We consider a set of one-dimensional classical oscillators, with equilibrium positions on the periodic $d$-dimensional lattice $\Z_N^d$. 
The phase space consists thus in the set of points 
\begin{equation*}
(q,p) 
\; = \; 
(q_x,p_x)_{x\in\Z_N^d}
\; \in \; 
(\R^2)^{N^d}.
\end{equation*}
The Hamiltonian is written as
\begin{equation*}
H(q,p) 
\; = \;
\frac{1}{2} \sum_{x \in\Z_N^d} \big( p_x^2 + \omega_x^2 q_x^2 \big)
+ \epsilon \sum_{x \in\Z_N^d} U (q_x)
+ \frac{\epsilon}{2} \sum_{x\in\Z_N^d} \sum_{y : |x-y|_1 = 1} V(q_x - q_y)
\end{equation*}
with the two following definitions. 

First, the eigenfrequencies of the decoupled harmonic oscillators, $\omega = (\omega_x)_{x\in\Z_N^d}$,
form a sequence of independent and identically distributed random variables, with law $\ProbaStar$ independent of $N$.
Let $\MeanStar (\cdot)$ be the associated expectation.
While more general distributions could likely be considered, 
we will assume that $\ProbaStar$ admits a bounded density, so that in particular $\MeanStar (\omega_x^2 ) - \MeanStar (\omega_x)^2  >  0$, and that, 
for some constants $0 < \omega_- < \omega_+ < + \infty$, the bound $\omega_- \le  \omega_x \le \omega_+$ holds almost surely. 

Second, 
the coupling constant $\epsilon$ is strictly positive.
The potentials $U$ and $V$ belong to $\mathcal C^\infty_{temp} (\R)$, the space of infinitely differentiable functions with polynomial growth. 
The pinning potential $U$ is taken such that, for every $\alpha > 0$, for some constant $c > 0$, and for all $x\in\R$, 
\begin{equation*}
\int_\R \ed^{-\alpha ( \omega_- y^2 + \epsilon  U (y) )} \, \dd y \; < \; +\infty
\qquad \text{and} \qquad
\omega_- + \epsilon \, U'' (x) \; \ge \; c.
\end{equation*}
The interaction potential $V$ is supposed to be symmetric, meaning that $V(-x) = V(x)$ for all $x\in\R$.
Moreover, in order to ensure good decay properties of the Gibbs measure, we require the pinning to dominate over the interaction:
similarly to (2.3) in \cite{bod}, one asks that there exists a constant $\mathrm C < +\infty$ such that, for every $x,y\in \R$,
\begin{equation*}
V'' (x - y)^4
\; \le \; 
\mathrm C \,  \big( 1 + U'' (x)^2 \big) \big( 1 + U'' (y)^2 \big). 
\end{equation*}

We possibly perturb the hamiltonian dynamics by a stochastic noise that preserves the energy of each oscillator. 
Let $\epsilon' \ge 0$. 
The full generator of the dynamics writes
\begin{equation*}
L 
\; = \; 
A_{har} + \epsilon A_{anh} + \epsilon' S 
\; = \; 
A_{har} + \epsilon A_{anh}^{(0)} + \epsilon A_{anh}^{(1)} + \epsilon' S 
\end{equation*} 
with
\begin{align*}
A_{har} u 
\; & = \; 
\sum_{x\in\Z_N^d} \big( p_x \, \partial_{q_x} u - \omega_x^2 q_x \, \partial_{p_x} u \big), \\
A_{anh}^{(0)} u
\; & = \; 
- \sum_{x\in\Z_N^d} U' (q_x) \, \partial_{p_x} u, \\
A_{anh}^{(1)} u 
\; & = \; 
- \sum_{x\in\Z_N^d}  \sum_{y: |x-y|_1 = 1} V' (q_x - q_y)  \, \partial_{p_x} u , \\
S u 
\; & = \;
\sum_{x\in\Z_N^d} \big( u (\dots, -p_x, \dots ) - u (\dots , p_x , \dots) \big). 
\end{align*}
We denote by $\big( X_t^{(\epsilon,\epsilon')} \big)_{t \ge 0}$, or simply by $(X_t)_{t \ge 0}$, the Markov process on $(\R^2)^{N^d}$ generated by $L$.

Besides $\MeanStar(\cdot)$, we will consider two other expectations. 
Let $0 < \beta < +\infty$ be an inverse temperature, that will be considered as a fixed parameter in the sequel.
Let $\langle \cdot \rangle_\beta$ be the corresponding Gibbs measure, defined by \eqref{density Gibbs measure} below.
We will just denote by $\Mean(\cdot)$ the expectation over the realizations of the noise. 
We will write $\Mean_\beta (\cdot)$ for $\langle \Mean (\cdot) \rangle_\beta$.

We need some extra informations on the Gibbs measure. 
Its density $\rho_\beta$ is given by 
\begin{equation*}\label{density Gibbs measure}
\rho_\beta (q,p) \; = \; \ed^{-\beta H(q,p)} / Z(\beta),
\end{equation*} 
where $Z(\beta)$ is a normalization constant.
This density factorizes as $\rho_\beta (q,p) \; = \; \rho'_\beta(q) \multiplication \rho''_\beta (p)$, 
and the density $\rho''_\beta$ factorizes again:
\begin{equation*}
\rho''_\beta (p) \; = \; \Big( \frac{\beta}{2\pi} \Big)^{N^d/2} \prod_{x\in\Z_N^d} \ed^{-\beta p_x^2 /2}.
\end{equation*}
While $\rho'_\beta(q)$ does not factorize, the conclusion of Theorem 3.2 in \cite{bod} holds thanks to our hypotheses on the hamiltonian $H$\footnote{
Strictly speaking, a slight adaptation of this theorem is needed to cover our needs: 
going through the proof, one checks that the randomness on the pinning does not affect the result, 
and that the constants can be taken independent of $\epsilon$.
}:
there exist constants $\mathrm C < + \infty$ and $c > 0$ such that, 
for any functions $u,v\in \mathcal C^{\infty}_{temp}(\R^{N^d})$ satisfying $\langle u \rangle_\beta = \langle v \rangle_\beta = 0$, 
it holds that
\begin{equation}\label{decorrelation Gibbs measure}
|\langle f\multiplication g \rangle_\beta| \; \le \; \mathrm C \, \ed^{-c d(S(u),S(v))} \langle (\nabla_q u)^2 \rangle_\beta^{1/2} \langle (\nabla_q v)^2 \rangle_\beta^{1/2}.
\end{equation}
Here, $S(u)$ is the support of a function $u$, defined as the smallest subset of $\Z_N^d$ such that $u$ can be
written as a function of the variables $q_x$ for $x \in S(u)$, whereas $d(S(u),S(u))$ is the smallest distance
between any point in $S(u)$ and any point in $S(v)$.

\section{Conductivity}\label{sec: Conductivity}

Let $x\in \Z_N^d$. The energy of particle $x$ is given by 
\begin{equation*}
e_x 
\; = \;
\frac{1}{2} (p_x^2 + \nu_x q_x^2) + \epsilon U (q_x) + \frac{\epsilon}{2} \sum_{y:|x-y|_1 = 1} V(q_x - q_y).  
\end{equation*}
For $y\in\Z_N^d$ such that $|x-y|_1 = 1$, one defines the current 
\begin{equation*}
\epsilon \, j_{x,y} \; = \; \epsilon \,\frac{p_x + p_y}{2} V'(q_x - q_y).
\end{equation*}
Since $V$ is symmetric, one has $j_{y,x} = - j_{x,y}$.
With this definition, one computes that 
\begin{equation*}
L e_x \; = \; \epsilon \, \sum_{k=1}^d \big( j_{x-\mathsf{e}_k,x} - j_{x,x+\mathsf{e}_k} \big)
\end{equation*}
where $\mathsf{e}_k$ is the $k^{\mathrm{th}}$ unit vector.

To simplify some further notations, we find it convenient to define now a set of oriented bonds in the lattice $\Z_N^d$: 
\begin{equation*}
\widetilde{\Z}_N^d \; = \;  \big\{ (x,y)\in (\Z_N^d)^2 : y = x + \mathsf{e}_k \text{ for some } 1 \le k \le d \big\}.
\end{equation*}
The total current and the rescaled total current are defined by 
\begin{equation}\label{total current}
\epsilon J_N \; = \; \epsilon \, \sum_{\widetilde{x}\in\widetilde{\Z}_N^d} j_{\widetilde{x}} 
\qquad \text{and} \qquad
\epsilon \mathcal J_N \; = \; \frac{\epsilon}{N^{d/2}} J_N.
\end{equation}
Then, if the limits exist, the Green-Kubo conductivity of the chain is defined by (see for example \cite{lep}) 
\begin{equation}\label{defintion Green-Kubo}
\kappa (\epsilon, \epsilon') 
\; = \;
\lim_{t\rightarrow \infty}\lim_{N\rightarrow \infty} \kappa(\epsilon,\epsilon',N,t)
\; = \; 
\lim_{t\rightarrow \infty}\lim_{N\rightarrow \infty} d^{-1}\beta^2 \, \Mean_\beta \bigg( \frac{\epsilon}{\sqrt t}\int_0^t \mathcal J_N \circ X_s^{(\epsilon,\epsilon')} \, \dd s \bigg)^2.
\end{equation}
It maybe could look more natural to define the total current, and the associated Green-Kubo conductivity, 
with respect to a given direction, say $\mathsf e_k$, for some $1 \le k \le d$.
This means that the sum in \eqref{total current} should be restricted to currents on bonds of the type $\widetilde{x} = (x,x+\mathsf e _k)$ for $x\in\Z_d^N$.
Our results below still hold with such a restriction. 

We first prove a result on the fall-off of the conductivity in a weak coupling limit ; 
it is physically most meaningful for $1 << m << n$:
\begin{Theorem}\label{the: faster than any power law}
Let $n\ge 1$ and let $\epsilon' = 0$. 
Let also $1 \le m \le n$.
For almost all realizations of the eigenfrequencies, it holds that 
\begin{equation*}
\lim_{t\rightarrow \infty} \limsup_{\epsilon\rightarrow 0} \limsup_{N\rightarrow \infty} \epsilon^{-m}
\bigg\langle \bigg( \frac{\epsilon}{\sqrt{\epsilon^{-n}t}}\int_0^{\epsilon^{-n}t} \mathcal J_N \circ X_s^{(\epsilon,0)} \, \dd s \bigg)^2 \bigg\rangle_\beta
\; = \; 0.
\end{equation*}
\end{Theorem}
If the energy conserving noise is added, the following also holds:  
\begin{Theorem}\label{the: faster than any power law bis}
Let $n\ge 1$ and assume that 
\begin{equation*}
\epsilon ' \; = \; \epsilon^n.
\end{equation*}
Then there exist constants $\mathrm C = \mathrm C (n) < +\infty$ and $\epsilon_0 > 0$ such that, 
for almost all realizations of the eigenfrequencies, and for all $\epsilon \in [0,\epsilon_0]$, one has
\begin{equation*}
\limsup_{t\rightarrow \infty}
\limsup_{N\rightarrow \infty}
\Mean_\beta \bigg( \frac{\epsilon}{\sqrt t}\int_0^t \mathcal J_N \circ X_s^{(\epsilon,\epsilon')} \, \dd s \bigg)^2
\; \le \; 
\mathrm C(n) \multiplication \epsilon^{n+2} .
\end{equation*}
\end{Theorem}

\Remarks
1. The hypothesis that the law of the frequencies $(\omega_x)_{x\in\Z_{N}^d}$ admits a density cannot be completely dropped out for the proof to work. 
Assume indeed, at the opposite, that the frequencies should take only two values.
In that case, among three oscillators, two at least have the same frequency.  
Then, as long as one takes $d = 1$ and $n > 1$, or $d\ge 2$, some fatal resonances appear and invalidate our line of reasoning.

2. It is possible that the weak coupling limit or the noise actually destroys some effects, such as a very slow ballistic transport,
that become dominant when looking to the system at large enough time-scales.
For example, let us assume that $n$ is given, that $d=1$, that all the interactions are harmonic, 
and that $(\omega_x)_{x\in\Z_N}$ form a sequence of $2n$ different frequencies, arranged periodically. 
One checks that our proof works in this particular case, although the true conductivity is known to be infinite.
Since however, in our theorems, $n$ can be taken arbitrarily large, such a scenario only becomes possible at time-scales larger than any inverse polynomial, 
and does not affect our main message that the transport of energy has a non-perturbative origin.  
 
3. The proofs
are achieved by gathering three different ideas.
Given $\widetilde{x}\in\widetilde{\Z}^N_d$, we first approximately solve the Poisson equation $-L u = j_{\widetilde{x}}$, 
assuming a non-resonance condition on the random frequencies of the oscillators located near $\widetilde{x}$, as defined in the beginning of the next Section. 
We next exploit the conservation of energy to see that the currents $j_{\widetilde{x}}$,
at places $\widetilde{x}$ such that this condition is violated, 
in fact do not contribute to the evaluation of the conductivity. 
We finally take into account that $\epsilon \rightarrow 0$ before that $t \rightarrow\infty$  for Theorem \ref{the: faster than any power law}, 
or that $S$ generates a diffusion on time scales of order $\epsilon^n$ for Theorem \ref{the: faster than any power law bis}, 
in order to get a bound on rest terms.

\section{Proof of Theorems \ref{the: faster than any power law} and \ref{the: faster than any power law bis}}\label{sec: Proof of Theorem}

The proofs of our two results are analogous but, since that of Theorem \ref{the: faster than any power law bis} is slightly more involved, we focus on it. 
The needed adaptations to show Theorem \ref{the: faster than any power law} are presented in the last Subsection. 

Let us first define what are non-resonant frequencies near a bond $\widetilde{x}\in\widetilde{\Z}^N_d$. 
Let $\xi = (\xi_x)_{x\in\Z_N^d}$ with $\xi_x \in\Z$ for all $x\in\Z_N^d$.
Given an integer $r \ge 0$, given $x\in\Z_N^d$ and given $\widetilde{x}=(a,b)\in\widetilde{\Z}_N^d$, we define
\begin{equation*}
\mathrm B_1 (x, r) \; = \; \{ y \in \Z_N^d : |y-x|_1 \le r \}
\qquad \text{and} \qquad
\mathrm B_1 (\widetilde{x},r) \; = \; \mathrm B_1 (a, r) \cup \mathrm B_1 (b, r).
\end{equation*}
Given $\widetilde{x}\in\widetilde{\Z}_N^d$, given a realization $\omega$ of the frequencies, and given an integer $r \ge 0$, we define
\begin{equation*}
\langle \omega , \xi \rangle_{\widetilde{x},r} \; = \; \sum_{y \in \mathrm B_1 (\widetilde{x},r)} \omega_y \xi_y.
\end{equation*}
For $\alpha,c \in \R$, for $\widetilde{x}\in\widetilde{\Z}_N^d$, and for an integer $r \ge 0$, we next define the diophantine set
\begin{equation*}
D_{\alpha,c}(\widetilde{x},r) \; = \; 
\Big\{ 
\omega : |\langle \omega , \xi \rangle_{\widetilde{x},r}| \ge \frac{c}{|\xi|_1^\alpha} 
\text{ for all } 
\xi \ne 0 
\text{ on }
\mathrm B_1(\widetilde{x},r)
\Big\} .
\end{equation*}
Here, $\xi \ne 0 \text{ on }\mathrm B_1(\widetilde{x},r)$ means that $\xi_x \ne 0$ for at least one $x\in\mathrm B_1(\widetilde{x},r)$.
The following monotony property holds: if $r\le r'$, then $D_{\alpha,c}(\widetilde{x},r) \subset D_{\alpha,c}(\widetilde{x},r')$.
It is known (see for example \cite{pos2}) that, given $\delta > 0$ and an integer $r \ge 0$, 
there exist constants $\alpha< +\infty$ large enough and $c>0$ small enough such that, for all $\widetilde{x}\in\widetilde{\Z}_N^d$, it holds that
\begin{equation}\label{measure non resonant set}
\ProbaStar \big( \omega \in D_{\alpha,c}(\widetilde{x},r) \big) \; \ge \; 1-\delta.
\end{equation}

From now, let $n\ge 1$ be as in the hypotheses.
Let $\delta > 0$ to be fixed in Subsection \ref{subsec: Resonances}, and let $\alpha, c> 0$ be such that \eqref{measure non resonant set} holds with $r = n$.
We also fix a realization $\omega$ of the frequencies. 
All the constants introduced below may depend on $n,\delta, \alpha$ and $c$.

\subsection{Approximate solution to the Poisson equation}\label{subsec: Approximate solution to the Poisson equation}

Let us start by a simple, but crucial, observation. 
We say that a function $u = u(q,p)$ on the phase space is $p$-symmetric if $u(q,p) = u(q,-p)$ for all $(q,p)\in(\R^{2})^{N^d}$ ; 
we similarly define $p$-antisymmetric functions. 
The generators $A_{har}$ and $A_{anh}$ map $p$-symmetric functions to $p$-antisymmetric functions, 
and $p$-antisymmetric functions to $p$-symmetric functions.  
The interplay between $p$-symmetric and $p$-antisymmetric functions has shown to have deep consequences in an other context \cite{ben}.

For the rest of this Subsection, we fix $\widetilde{x}\in\Z_N^d$, and we assume that $\omega \in D_{\alpha, c}(\widetilde{x},n)$.
For an integer $k \ge 0$, we designate by $||| \cdot |||_k$ the norm of the Sobolev space $\mathrm H^{k}(\R^{2N^d},\langle \cdot \rangle_\beta)$ ;
explicitly
\begin{equation*}
||| u |||_k \; = \; \sum_{\alpha \in \N^{2N^d} : |\alpha|_1 \le k} || \partial^\alpha u ||_{\mathrm L^2 (\R^{2N^d},\langle \cdot \rangle_\beta)}
\quad \text{with} \quad 
|\alpha|_1 = \sum_{j=1}^{2N^d}\alpha_j \quad \text{and} \quad \partial^\alpha u = \partial^{\alpha_1}_{x_1} \dots \partial^{\alpha^{2N^d}}_{x_{2N^d}} .
\end{equation*}
Let us first show the two following lemmas. 

\begin{Lemma}\label{lem: Poisson equation harmonic}
Let $\omega \in D_{\alpha, c}(\widetilde{x},n)$.
Let $f\in \mathcal C^{\infty}_{temp}(\R^{2N^d})$ be a $p$-antisymmetric function
that depends only on the variables $(q_y,p_y)$ with $y\in\mathrm B_1 (\widetilde{x},r)$ for some $0 \le r \le n-1$. 
Then there exists a $p$-symmetric function $u \in \mathcal C^{\infty}_{temp}(\R^{2N^d})$,
depending on the same variables as $f$, 
satisfying $\langle u \rangle_\beta = 0$,
and solving 
\begin{equation*}
-A_{har} u \; = \; f.
\end{equation*}
Moreover, there exists a integer $m\ge 0$ such that, for any $k\ge 0$ and for some constant $\mathrm C_k < +\infty$, 
\begin{equation*}
||| u |||_k \; \le \; \mathrm C_k \, ||| f |||_{k+m}.
\end{equation*}
\end{Lemma}

\noindent
It follows from this Lemma that $A_{anh} u \in \mathcal C^{\infty}_{temp}(\R^{2N^d})$
is $p$-antisymmetric, and depends only on the variables $(q_y,p_y)$ with $y\in\mathrm B_1 (\widetilde{x},r+1)$.
Moreover, there exists some non-negative integer $m'$ such that, for any $k\ge 0$ and for some constant $\mathrm C_k < +\infty$, 
\begin{equation*}
||| A_{anh} u |||_k \; \le \; \mathrm C_k \, ||| f |||_{k+m'}.
\end{equation*}

\Proof
For $z>0$, the solution $u_z\in \mathcal C_{temp}^\infty(\R^{2N^d})$ to the equation
\begin{equation*}
(z - A_{har}) u_z \; = \; f
\end{equation*}
exists and is unique ; it is given by 
\begin{equation}\label{abstract definition of u z}
u_z 
\; = \; 
\int_0^\infty \ed^{-zt} \ed^{A_{har} t} f \, \dd t 
\; = \;
\int_0^\infty \ed^{-zt} f \circ X_t^{har} \, \dd t
\end{equation}
where $(X_t^{har})_{t\ge 0}$ is the deterministic process generated by $A_{har}$.
Since $f\in \mathcal C^\infty_{temp}(\R^{2N^d})$, and since $(X_t^{har})_{t\ge 0}$ explicitly given by \eqref{uncoupled process}, it is checked that $u_z$ defined by \eqref{abstract definition of u z}
is indeed an element of $\mathcal C_{temp}^\infty(\R^{2N^d})$. 
We will show that there exists a function $u$, 
having all the regularity properties appearing in the conclusions of the Lemma,
depending on the same variables as $f$, and such that 
\begin{equation}\label{to be satisfied first lemma}
\lim_{0 < z \rightarrow 0} u_z \; = \; u
\qquad \text{and} \qquad
\lim_{0 < z \rightarrow 0} A_{har} u_z \; = \; A_{har}u,
\end{equation}
these limits being for example understood in $\Lp^2(\R^{2N^d})$. 
This implies that $u$ solves the equation $-A_{har} u = f$. 
So does then the function $u - \langle u \rangle_\beta$, which inherits also from the other properties of $u$ mentioned in the conclusions of the Lemma. 
This will thus conclude the proof. 

For given initial conditions $(q,p)\in\R^{2N^d}$, one has 
\begin{equation}\label{uncoupled process}
X_t^{har} (q,p) 
\; = \;
\big( \dots , q_k \cos \omega_k t + (p_k/\omega_k) \sin \omega_k t , -\omega_k q_k \sin \omega_k t + p_k \cos \omega_k t , \dots  \big). 
\end{equation}
Therefore, defining
\begin{equation*}
g(q,p ; \theta) \; = \; 
f\big( \dots , q_k \cos \theta_k + (p_k/\omega_k) \sin \theta_k , -\omega_k q_k \sin \theta_k  + p_k \cos \theta_k  , \dots \big)
\end{equation*} 
and
\begin{equation*}
\theta (t) \; = \; \omega t \in \R^{N^d},  
\end{equation*}
one has
\begin{equation*}
u_z (q,p) \; = \; \int_0^\infty \ed^{-zt} g (q,p; \theta (t))\, \dd t.
\end{equation*}

It is convenient to write $g$ in Fourier's variables:
\begin{equation*}
g (q,p;\theta) \; = \; \sum_{\xi\in\Z^{N^d}} \hat{g}(q,p;\xi) \, \ed^{i \theta\cdot \xi}
\qquad \text{with} \qquad
\hat{g}(q,p;\xi) \; = \; \frac{1}{(2\pi)^{N^d}} \int_{[0,2\pi]^{N^d}} g (q,p;\theta)  \, \ed^{-i \theta\cdot \xi} \, \dd \theta. 
\end{equation*} 
Since $f$ only depends on the variables $(q_y,p_y)$ with $y \in \mathrm B (\widetilde{x},r)$, 
it holds that $\hat{g}(q,p,\xi) = 0$ for any $(q,p)$, as soon as $\xi_y \ne 0$ for some $y \notin \mathrm B(\widetilde{x},r)$. 
Let us see that $\hat{g}(q,p;0) = 0$ for any $(q,p)$. 
For this, let us define $\overline{\theta} = \overline{\theta}(q,p) \in [0,2\pi]^{N^d}$ by 
\begin{equation*}
q_k + ip_k /\omega_k \; = \; \big( q_k^2 + p_k^2 /\omega_k^2\big)^{1/2} \exp (i \overline{\theta}_k).
\end{equation*}
The $p$-antisymmetry of $f$ implies that
\begin{equation*}
g (q,p ; 2 \overline{\theta} - \theta) \; = \; - g (q,p;\theta), 
\end{equation*}
from which $\hat{g}(q,p;0) = 0$ follows.

One now computes that 
\begin{equation*}
u_z (q,p) 
\; = \;
\sum_{\xi \in\Z^{N^d} - \{ 0 \}} \hat{g} (q,p ; \xi) \int_0^{\infty} \ed^{-zt} \ed^{i t \omega \cdot \xi } \, \dd t
\; = \; 
\sum_{\xi \in\Z^{N^d} - \{ 0 \}} \frac{\hat{g} (q,p ; \xi)}{z - i \omega \cdot \xi}
\end{equation*}
and so, taking the limit $z\rightarrow 0$, 
\begin{equation*}
u (q,p) 
\; = \;
 i \sum_{\xi \in\Z^{N^d} - \{ 0 \}} \frac{\hat{g} (q,p ; \xi)}{\langle \omega \cdot \xi \rangle_{\widetilde{x},r}}.
\end{equation*}
The regularity properties of $u$ are obtained thanks to the hypothesis $\omega \in D_{\alpha,c}(\widetilde{x},n)$, 
as is usual in KAM theory ; see for example \cite{pos2}. From there, \eqref{to be satisfied first lemma} follows.
$\square$

\begin{Lemma}\label{lem: Poisson equation symmetric}
Let $f\in \mathcal C^{\infty}_{temp}(\R^{2N^d})$ be a $p$-antisymmetric function
that depends only on the variables $(q_y,p_y)$ with $y\in\mathrm B_1 (\widetilde{x},r)$ for some $0 \le r \le n-1$. 
Then there exists a function $u \in \mathcal C^{\infty}_{temp}(\R^{2N^d})$, satisfying $\langle u \rangle_\beta = 0$,
depending on the same variables as $f$, 
and solving 
\begin{equation*}
-S u \; = \; f.
\end{equation*}
Moreover, for any $k\ge 0$ and for some constant $\mathrm C_k < +\infty$, it holds that $||| u |||_k \; \le \; \mathrm C_k \, ||| f |||_{k}$.
\end{Lemma}

\Proof
Let us fix the values of $|p_x|$ and $q_x$ for all $x\in\Z_N^d$, 
so that in fact $f$ is now seen as a function on $\{-1,+1 \}^{|B_1(\widetilde{x},r)|}$, the set of signs of the impulsions.
The operator $S$ generates a standard continuous time random walk on this set. 
Writing  $S = |B_1(\widetilde{x},r)| (T - Id)$, one has $u = \frac{1}{|B_1(\widetilde{x},r)|} \sum_{k\ge 0} T^k f$.
The terms $T^k f$ converge exponentially fast to 0, provided that $f$ is of mean zero with respect to the uniform measure on $B_1(\widetilde{x},r)$.
So it is since $f$ is $p$-antisymmetric. The regularity of $u$ follows form this representation.
$\square$

The function $j_{\widetilde{x}}$ is $p$-antisymmetric, belongs to $\mathcal C^{\infty}_{temp}(\R^{2N^d})$, and depends only on the variables in $\mathrm B(\widetilde{x},0)$.
Lemma \ref{lem: Poisson equation harmonic} ensures that there exist functions $u^{(1)}_{\widetilde{x}} , \dots , u^{(n)}_{\widetilde{x}} $ solving the following finite hierarchy
\begin{align}
- A_{har} u^{(1)}_{\widetilde{x}} \; &= \; j_{\widetilde{x}}, \label{perturbation order 1}\\
- A_{har} u^{(2)}_{\widetilde{x}} \; &= \; A_{anh} u^{(1)}_{\widetilde{x}}, \label{perturbation order 2}\\
&\dots \nonumber\\
-A_{har} u^{(n)}_{\widetilde{x}} \; &= \; A_{anh} u^{(n-1)}_{\widetilde{x}}. \label{perturbation order n}
\end{align}
By Lemma \ref{lem: Poisson equation symmetric}, there exists also a function $v^{(n)}_{\widetilde{x}}$ such that 
\begin{equation*}
- S v^{(n)}_{\widetilde{x}} \; = \; A_{anh} u^{(n)}_{\widetilde{x}}.
\end{equation*}
Defining
\begin{equation*}
u_{\widetilde{x}} \; = \; u^{(1)}_{\widetilde{x}} + \epsilon u^{(2)}_{\widetilde{x}} + \dots + \epsilon^{n-1} u^{(n)}_{\widetilde{x}}
\qquad 
\text{and}
\qquad
v_{\widetilde{x}}  \; = \; u_{\widetilde{x}} - v^{(n)}_{\widetilde{x}}
\end{equation*}
one finds that 
\begin{equation}\label{approximate Poisson decomposition current}
j_{\widetilde{x}} \; = \; -L u_{\widetilde{x}} + \epsilon^n (A_{anh} u^{(n)}_{\widetilde{x}} + S u_{\widetilde{x}}) \; = \; -L u_{\widetilde{x}} + \epsilon^n S v_{\widetilde{x}}.
\end{equation}
The functions $u_{\widetilde{x}}$ and $ v_{\widetilde{x}}$ belong to $\mathcal C^{\infty}(\R^{2N^d})$, 
are of mean zero,
depend only on the variables $(q_y,p_y)$ with $y \in \mathrm B_{1}(\widetilde{x},n)$, 
and have (Sobolev) norms that do not depend on $\widetilde{x}$, as long as $\omega \in D_{\alpha,c}(\widetilde{x},n)$. 

\Remark
While it is not difficult to solve the finite hierarchy \eqref{perturbation order 1}-\eqref{perturbation order 2} as we did under a non-resonant assumption on $\omega$,
one could in fact have expected it to be much less straightforward.  
Indeed, the dynamics generated by $A_{har}$ is very degenerated since it preserves the energy of every single particle.
This has two consequences. 
First, the equation $-A_{har} u = f$ can only be solved if $f$ is of mean zero with respect to every microcanonical surface defined by the value of the energies of each uncoupled particle.
Second, if this is the case, then $u$ is not unique, since every function depending only on the energies of decoupled particles lies in the kernel of $A_{har}$.
The fact that the currents are $p$-antisymmetric, and that $A_{anh}$ interchanges $p$-symmetric and $p$-antisymmetric functions,
guarantees therefore that the zero-mean condition is satisfied at all orders, without that we have to select a particular solution at each step. 
The operator $S$ instead destroys this symmetry of the system, allowing diffusion of energy.

\subsection{Resonances}\label{subsec: Resonances}

We fix a realization $\omega$ of the frequencies. 
We let 
\begin{equation*}
A_N \; = \; \{ \widetilde{x} \in \widetilde{\Z}_N^d :  \omega \in D_{\alpha, c}(\widetilde{x},n)\} .
\end{equation*}
We then partition $\widetilde{\Z}_N^d - A_N$ as
\begin{equation*}
\widetilde{\Z}_N^d - A_N
\; = \; 
\bigcup_{\gamma} \mathcal C_\gamma,
\end{equation*} 
where $(\mathcal C_\gamma)_\gamma$ is the set of connected components, or clusters, of $\widetilde{\Z}_N^d - A_N$.
We let $\ell(\mathcal C_\gamma)\ge 1$ be the length of the longest self-avoiding path in $\mathcal C_\gamma$.
Given $\widetilde{x}\in\widetilde{\Z}_N^d - A_N$, 
we also define $\mathcal C(\widetilde{x})$ as the cluster $\mathcal C_\gamma$ such that $\widetilde{x}\in \mathcal C_\gamma$.

In the sequel, by a slight abuse of notation, we will say that $x\in\Z_N^d$ belongs to $E \subset \widetilde{\Z}_N^d$, 
if there exists a bond $\widetilde{x}\in E$ such that $x$ is a component of $\widetilde{x}$.
So any subset $E\subset \widetilde{\Z}_N^d$ will also be considered as a subset of $\Z^d_N$.
Given $\mathcal K \subset \Z_N^d$, we define its border $\partial \mathcal K \subset \widetilde{\Z}_N^d$ as
as the set of bonds $\widetilde{x} = (a,b)$ such that
\begin{equation*}
\partial \mathcal K
\; = \;
\big\{ 
\widetilde{x} = (a,b) \in \widetilde{\Z}_N^d :
(a \in \mathcal K \text{ and }  b\notin \mathcal K )
\quad \text{or} \quad
(a \notin \mathcal K \text{ and } b\in \mathcal K )
\big\} . 
\end{equation*}
Given $\mathcal K \subset \widetilde{\Z}_N^d$, one defines $\partial \mathcal K$ as the border of $\mathcal K$ seen as a subset of $\widetilde{\Z}_N^d$.
The following Lemma contains the core of the argument that allows us to get rid of resonances.
\begin{Lemma}\label{lem: conservation of energy}
Let $\mathcal C_\gamma \subset \widetilde{\Z}_N^d - A_N$ be a cluster, 
and let $\ell (\mathcal C_\gamma ) = m$.  It holds that 
\begin{equation*}
\sum_{\widetilde{x}\in\mathcal C_\gamma} j_{\widetilde{x}}
\; = \; 
\sum_{\widetilde{x}\in \partial\mathcal C_\gamma} \theta_1 (\widetilde{x},\gamma) \, j_{\widetilde{x}}
\, - \, 
\epsilon^{-1} L \Big( \sum_{x\in \mathcal C_\gamma \subset \Z_N^d} \theta_2 (x) \, e_x \Big)
\end{equation*}
where $\theta_1$ and $\theta_2$ are integer valued functions that satisfy the bounds
\begin{equation*}
\sup_{\widetilde{x} \in \partial\mathcal C_\gamma} |\theta_1 (\widetilde{x}, \gamma)| \; \le \; \mathrm C \, m^d, 
\qquad
\sup_{x \in \mathcal C_\gamma} |\theta_2 (x)| \; \le \; \mathrm C \, m^{d},
\end{equation*}
for a constant $\mathrm C$ that depends only on the dimension $d$.
\end{Lemma}

\begin{figure}[h]
\begin{center}
\includegraphics[width=0.48\textwidth]{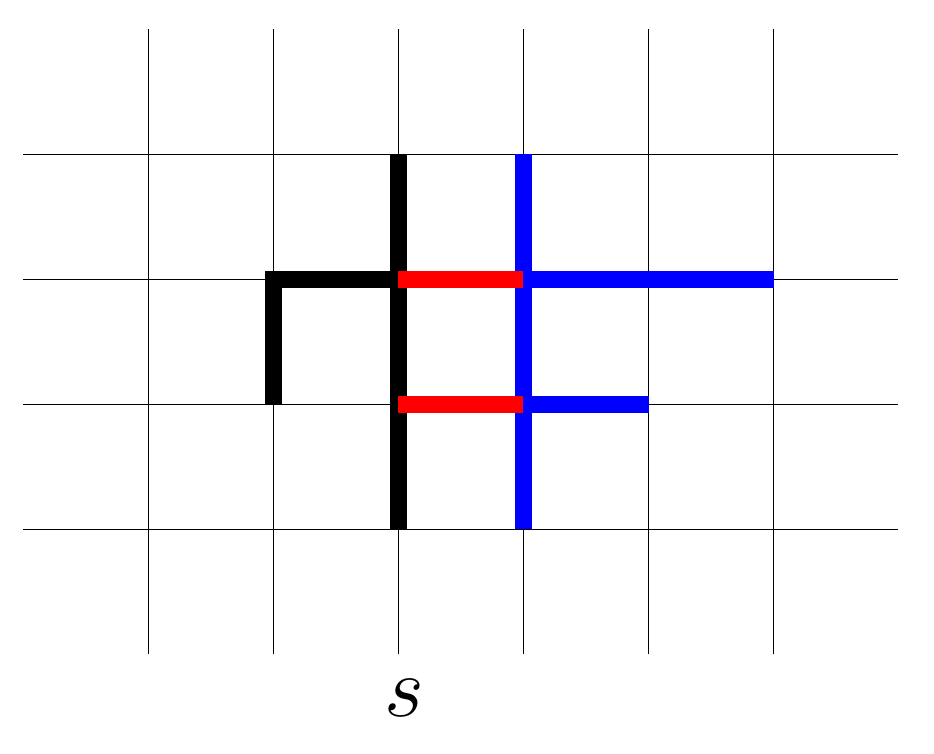}
\end{center}

\vspace{-0.8cm}

\caption{\label{figure: resonant cluster}
The sets $\mathcal C_\gamma$, $P(s,k)$ and $\mathcal K (s,k)$ appearing in the proof of Lemma \ref{lem: conservation of energy}.
Here $d=2$ and $k=1$. The whole figure represent a cluster $\mathcal C_\gamma$. 
For a given $s$, the bonds in red constitute the set $P(s,k=1)$, while the bonds in blue constitute the set $\mathcal K (s,k=1)$. 
}
\end{figure}

\Proof 
Given $\mathcal K \subset\Z_N^d$, the conservation of energy implies
\begin{equation}\label{divergence theorem}
\epsilon^{-1} L \bigg(  \sum_{x \in \mathcal K} e_x \bigg) \; = \; \sum_{\widetilde{x}\in \partial \mathcal K} \zeta (\widetilde{x}) \, j_{\widetilde{x}}.
\end{equation}
Here, writing $\widetilde{x} = (a,b)$, one has $\zeta (\widetilde{x}) = 1$ if $b\in\mathcal K$ and $\zeta (\widetilde{x}) = -1$ if $a\in\mathcal K$.
To apply this formula, 
we can repeatedly consider contours that partition $\mathcal C_\gamma$ into two pieces, and get rid of all the currents in $\mathcal C_\gamma$.
A systematical way to do this is as follows (see Figure \ref{figure: resonant cluster}).
Let us give a direction $k\in \{1, \dots ,d \}$ and a value $s\in \Z_N$ 
such that for some $\widetilde{x} = (a,a+\mathsf{e}_k)\in\mathcal C_\gamma$, it holds that $a_k =s$, where $a_k$ denotes the $k^{\mathrm{th}}$ coordinate of $a$. 
We then define 
\begin{equation*}
P (s,k) \; = \; \{ \widetilde{x}=(a,a+\mathsf e_k)\in \mathcal C_\gamma : a_k = s \} ,
\end{equation*}
as well as the set $\mathcal K(s,k)\subset \mathcal C_\gamma \subset \Z_N^d$ made of points $x$ such that $x_k > s$.
Then, using \eqref{divergence theorem},
\begin{equation*}
\epsilon^{-1} L\bigg( \sum_{x \in \mathcal K(s,k)} e_{x} \bigg) 
\; = \; 
\sum_{\widetilde{x}\in P (s,k) } j_{\widetilde{x}} 
+ 
\sum_{\widetilde{x}\in \partial \mathcal K(s,k) - P(s,k)} \zeta (\widetilde{x}) \, j_{\widetilde{x}}.
\end{equation*}
It hols that $\partial \mathcal K(s,k) - P(s,k) \subset \partial \mathcal C_\gamma$.
This operation can be repeated with every direction $k\in \{1, \dots ,d \}$ and every value  $s\in \Z_N$. 
This only needs to be done $\mathcal O (m^d)$ times to cover the whole cluster ; summing one obtains
\begin{equation*}
-\epsilon^{-1} L\bigg( \sum_{x \in \mathcal C_\gamma} \theta_2 (x) e_{x} \bigg) 
\; = \; 
\sum_{\widetilde{x}\in \mathcal C_\gamma } j_{\widetilde{x}}
\,  - 
\sum_{\widetilde{x}\in\partial\mathcal C_\gamma} \theta_1 (\widetilde{x},\gamma) \, j_{\widetilde{x}} ,
\end{equation*}
for some functions $\theta_1$ and $\theta_2$ that satisfy the stated bounds.
$\square$

Thanks to this Lemma, and using the fact that $L \langle e \rangle_\beta = 0$, we decompose the total current as follows
\begin{align}
J_N 
\; & = \; 
\sum_{\widetilde{x}\in A_N} j_{\widetilde{x}} \, + \, \sum_{\gamma} \Big( \sum_{\widetilde{x}\in\mathcal C_\gamma} j_{\widetilde{x}} \Big)
\; = \; 
\sum_{\widetilde{x}\in A_N} j_{\widetilde{x}} 
\, + \, \sum_{\gamma} \Big( \sum_{\widetilde{x}\in \partial\mathcal C_\gamma} \theta_1 (\widetilde{x}) \, j_{\widetilde{x}} \Big)
\, - \, \epsilon^{-1} L\bigg( \sum_{\gamma} \Big( \sum_{x\in \mathcal C_\gamma} \theta_2 (x) \, e_x \Big) \bigg) 
\nonumber\\
\; & =: \;
 \sum_{\widetilde{x}\in A_N} \psi_1 (\widetilde{x}) \, j_{\widetilde{x}} 
\, - \, \epsilon^{-1} L\bigg( \sum_{x \in \Z_N^d} \psi_2 (x) \, \big( e_x - \langle e \rangle_\beta \big) \bigg).
\label{decomposition current resonances}
\end{align}
Here 
\begin{equation*}
\psi_1 (\widetilde{x}) \; = \; 1 + \sum_{\gamma: \widetilde{x} \in \partial \mathcal C_\gamma} \theta_1 (\widetilde{x},\gamma),
\end{equation*}
the sum containing two non-zero terms at most, and
\begin{equation*}
\psi_2 (x) \; = \; \theta_2 (x) \;\;\text{if} \;\; x \in \bigcup_{\gamma} \mathcal C_\gamma
\qquad \text{and} \qquad
\psi_2(x) \; = \; 0 \;\;\text{otherwise}.
\end{equation*}
%
%
The functions $\psi_1$ and $\psi_2$ inherit the properties of $\theta_1$ and $\theta_2$ stated in Lemma \ref{lem: conservation of energy}.
If $\widetilde{x}\notin \bigcup_{\gamma}\partial \mathcal C_\gamma$, then $|\psi_1 (\widetilde{x})| = 1$, 
whereas if $\widetilde{x}\in \bigcup_{\gamma}\partial \mathcal C_\gamma$, then $|\psi_1 (\widetilde{x})| \le \mathrm C \, m^d$, 
where $m= \ell (\mathcal C_\gamma)$, where $\mathcal C_\gamma$ is the largest cluster such that $\widetilde{x}\in \partial\mathcal C_\gamma$.
Similarly, if $x\notin \bigcup_{\gamma} \mathcal C_\gamma$, then $|\psi_2 (x)| = 0$, 
while if $x\in \mathcal C_\gamma$ for some $\gamma$, then $|\psi_2 (x)| \le \mathrm C \, m^d$, with $m = \ell (\mathcal C_\gamma)$. 

We now may fix the parameter $\delta$ introduced at the beginning of this Section. 
We take $\delta>0$  small enough so that there exist constants $\mathrm C < + \infty$ and $c' > 0$ such that,
for every integer $m \ge 1$, and for every $\widetilde{x}\in\Z_N^d$, it holds
\begin{equation}\label{proba cluster of size m}
\Proba_\star \big( \ell (\mathcal C(\widetilde{x})) = m \big) \; \le \; \mathrm C \, \ed^{-c' m}. 
\end{equation}
The number $\delta$ is well defined. 
To see this, we first notice that, for $\widetilde{x},\widetilde{y}\in\widetilde{\Z}_N^d$,
the events $\omega \notin D_{\alpha,c}(\widetilde{x},n)$  and $\omega \notin D_{\alpha,c}(\widetilde{y},n)$ become independent 
as soon as $\mathrm B_1 (\widetilde{x},n) \cap \mathrm B_1 (\widetilde{y},n) = 0$. 
Next, we observe that the number of self-avoiding paths in $\widetilde{\Z}_N^d$ of length $m$ is bounded by $(2d-1)^m$.
But the number of disjoints balls  of fixed radius $n$ in $\widetilde{\Z}_N^d$, with center contained in any such path, is proportional to $m$ as $m\rightarrow \infty$.
Therefore, the probability of forming a path of length $m$ in $\widetilde{\Z}_N^d - A_N$ decays exponentially with $m$ if $\delta$ is small enough.

\begin{Lemma}
For every $\widetilde{x}\in\widetilde{\Z}_N^d$, we let $f_{\widetilde{x}}$ be a smooth function on the phase space $\R^{2N^d}$, 
that satisfies $\langle f_{\widetilde{x}} \rangle_\beta = 0$,
that depends only on the variables in $\mathrm B (\widetilde{x}, n)$, 
and which $\mathrm H^1$-norm can be bounded by a constant independent of $\widetilde{x}$.
Then, there exists a constant $\mathrm C < +\infty$, such that, almost surely,
\begin{align}
\limsup_{N\rightarrow \infty} \bigg\langle  \Big(
\frac{1}{N^{d/2}} \sum_{\widetilde{x}\in A_N} \psi_1 (\widetilde{x}) \, f_{\widetilde{x}} 
\Big)^2\bigg\rangle_\beta
\; &\le \; 
\mathrm C, \label{uniform bound psi 1}\\
\limsup_{N\rightarrow \infty}\bigg\langle 
\bigg( \frac{1}{N^{d/2}}\sum_{x\in \Z_N^d} \psi_2 (x) \, \big( e_x - \langle e \rangle_\beta \big) \bigg)^2
\bigg\rangle_\beta
\; & \le \; 
\mathrm C. \label{uniform bound psi 2}
\end{align}
\end{Lemma}

\Proof
Throughout the proof of this lemma, $\mathrm C < + \infty$ and $c' > 0$ denote constants.
The proof of both bounds are analogous, and we only handle the first one.
Expanding the square and applying Cauchy-Schwarz inequality yields
\begin{align*}
V_N 
\; := \;
 \bigg\langle  \Big(
\frac{1}{N^{d/2}} \sum_{\widetilde{x}\in A_N} \psi_1 (\widetilde{x}) \, f_{\widetilde{x}} 
\Big)^2\bigg\rangle_\beta
\; \le \; 
\frac{1}{N^d} \sum_{\widetilde{x},\widetilde{y} \in A_N} \psi_1 (\widetilde{x}) \psi_1 (\widetilde{y})\, | \langle  f_{\widetilde{x}}   f_{\widetilde{y}} \rangle_\beta | 
\; \le \; 
\frac{1}{N^d} \sum_{\widetilde{x},\widetilde{y} \in A_N} \psi_1^2 (\widetilde{x}) \, | \langle  f_{\widetilde{x}}   f_{\widetilde{y}} \rangle_\beta | 
\end{align*} 
Then, first, thanks to the bounds on $|\psi_1 (\widetilde{x})|$ stated after the proof of Lemma \ref{lem: conservation of energy}, 
and thanks to the exponential bound \eqref{proba cluster of size m} on the probability of forming clusters of size $m$, 
it holds that 
\begin{equation}\label{norme L p}
\Mean_\star \big(|\psi_1 (\widetilde{x}) |^p \big) \le \mathrm C_p, 
\end{equation}
for any $p\ge 1$, and for some constant $\mathrm C_p < +\infty$ independent of $\widetilde{x}\in\widetilde{\Z}_N^d$. 
Second, thanks to the decorrelation bound \eqref{decorrelation Gibbs measure} on the Gibbs measure, and to the hypotheses on the functions $f_{\widetilde{x}}$, 
it holds that 
\begin{equation*}
\rho_{\widetilde x} \; := \; \sum_{\widetilde{y}\in A_N}  | \langle  f_{\widetilde{x}}   f_{\widetilde{y}} \rangle_\beta | \; \le \; \mathrm C.
\end{equation*}
Therefore
\begin{equation*}
V_N \; \le \; 
\frac{1}{N^d} \sum_{\widetilde{x}\in A_N} \big( \psi_1^2 (\widetilde{x}) - \Mean_\star (\psi_1^2 (\widetilde{x})) \big) \, \rho_{\widetilde{x}}
+ 
\frac{1}{N^d} \sum_{\widetilde{x}\in A_N} \Mean_\star (\psi_1^2 (\widetilde{x})) \, \rho_{\widetilde{x}}
\; = : \; 
Y_N + Z_N.
\end{equation*}

Since $Z_N$ is a bounded deterministic sequence, it suffices now to bound $Y_N$. 
By the Borel-Cantelli lemma and Markov inequality, the proof will be concluded if one finds $p\ge 1$ such that
\begin{equation}\label{Borel Cantelli Markov}
\sum_{N\ge 1} \Mean_\star \big( |Y_N|^p \big) \; < \; + \infty.
\end{equation} 
If $d = 1$, one can take $p=4$, but if $d \ge 2$, the choice $p=2$ suffices.
We only will deal with this second case ; the case $d=1$ is obtained similarly at the price of slightly longer computations.
To simplify some further notations, let us write
\begin{equation*}
\phi (\widetilde{x}) \; = \; \big( \psi_1^2 (\widetilde{x}) - \Mean_\star (\psi_1^2 (\widetilde{x})) \big).
\end{equation*}
One has 
\begin{equation}\label{sum for p = 2}
\Mean_\star \big( Y_N^2 \big)
\; \le \; 
\frac{\mathrm C}{N^{2d}} \sum_{\widetilde{x},\widetilde{y} \in A_N} \Mean_\star \big( \phi (\widetilde{x}) \phi (\widetilde{y}) \big).
\end{equation}

For $\widetilde{x},\widetilde{y}\in \widetilde{\Z}_N^d$, let us set 
$|\widetilde{x} - \widetilde{y}|_1
 = 
\min \{ |x- y|_1 : x\in \widetilde{x}, y \in \widetilde{y} \}.$
By \eqref{norme L p}, the terms in the sum in the right hand side of \eqref{sum for p = 2} are uniformly bounded, 
and it now suffices to establish decay bounds on off-diagonal terms.
Let us fix some $\widetilde{x},\widetilde{y} \in A_N$.
The claims below are valid for $|\widetilde{x} - \widetilde{y}|_1$ large enough.
Given $c'' > 0$, we then define an event $E$ as follows: 
$\omega\in E$ if nor $\widetilde{x}$ nor $\widetilde{y}$ do belong to the border of a cluster of size larger or equal to $c'' |\widetilde{x}-\widetilde{y}|$.
Then, if $c''$ small enough, 
the variables $\phi(\widetilde{x})$ and $\phi(\widetilde{y})$ become independent with respect to $\Proba_\star (\cdot | E)$.
Therefore
\begin{align*}
\Mean_\star \big( \phi(\widetilde{x})\phi(\widetilde{y}) \big)
\; &= \;
 \Mean_\star \big( \phi(\widetilde{x})\phi(\widetilde{y}) \big| E\big) \Proba_\star (E) 
\, + \,
\Mean_\star \big( \phi(\widetilde{x})\phi(\widetilde{y}) \big| E^{\mathsf c}\big) \Proba_\star (E^{\mathsf c}) \\
\; & =  \; 
\Mean_\star \big( \phi(\widetilde{x})\big| E\big) \Mean_\star \big( \phi(\widetilde{y})\big| E\big) \Proba_\star (E) 
\, + \,
\Mean_\star \big( \phi(\widetilde{x})\phi(\widetilde{y}) \big| E^{\mathsf c}\big) \Proba_\star (E^{\mathsf c}) \\
\; & =  \; 
- \, 
\Mean_\star \big( \phi(\widetilde{x})\big| E\big) \Mean_\star \big( \phi(\widetilde{y})\big| E^{\mathsf c}\big) \Proba_\star (E^{\mathsf c}) 
\, + \,
\Mean_\star \big( \phi(\widetilde{x})\phi(\widetilde{y}) \big| E^{\mathsf c}\big) \Proba_\star (E^{\mathsf c}),
\end{align*}
where one has used the fact that $\Mean_\star (\phi (\widetilde{y})) = 0$ to get the last line.
The exponential bound $\Proba_\star (E^{\mathsf c}) \le \mathrm C \,\ed^{-c' |\widetilde{x} - \widetilde{y}|_1}$ holds, 
so that in particular $\Proba_\star (E)$ is bounded form below by a strictly positive constant. So it holds that 
\begin{align*}
|\Mean_\star \big( \phi(\widetilde{x})\big| E\big)| 
\; &\le \; 
\mathrm C |\Mean_\star \big( \phi(\widetilde{x})\big)| \; \le \; \mathrm C, \\
|\Mean_\star \big( \phi(\widetilde{y})\big| E^{\mathsf c}\big) \Proba_\star (E^{\mathsf c}) |
\; &\le \;
\Mean_\star \big( \phi^2(\widetilde{y})\big)^{1/2} \Proba_\star^{1/2}(E^{\mathsf c}) \; \le \; \mathrm C\, \ed^{-c' |\widetilde{x} - \widetilde{y}|_1}, \\
|\Mean_\star \big( \phi(\widetilde{x})\phi(\widetilde{y}) \big| E^{\mathsf c}\big) \Proba_\star (E^{\mathsf c})|
\; &\le \; 
\Mean_\star \big( \phi^2(\widetilde{x})\phi^2(\widetilde{y})\big)^{1/2} \Proba_\star^{1/2}(E^{\mathsf c}) \; \le \; \mathrm C\, \ed^{-c' |\widetilde{x} - \widetilde{y}|_1}.
\end{align*}
Thus
\begin{equation*}
|\Mean_\star \big( \phi(\widetilde{x})\phi(\widetilde{y}) \big)| \; \le \; \mathrm C \, \ed^{-c' |\widetilde{x} - \widetilde{y}|_1}.
\end{equation*}
Inserting this estimate in \eqref{sum for p = 2}, it follows that $\Mean_\star (Y_N^2) \le \mathrm C/N^d$.
Since we have assumed $d\ge 2$, this in turn yields \eqref{Borel Cantelli Markov}, and so concludes the proof.
$\square$

\subsection{Concluding the proof of Theorem \ref{the: faster than any power law bis}}

We fix a realization $\omega$ of frequencies. 
We combine \eqref{approximate Poisson decomposition current} and \eqref{decomposition current resonances} to get
\begin{equation*}
J_N
\; = \; - \, 
L  \bigg( \sum_{\widetilde{x}\in A_N} \hspace{-0.1cm} \psi_1(\widetilde{x})\, u_{\widetilde{x}}  \bigg)
\, -\,
\epsilon^{-1} L \bigg( \sum_{\widetilde{x}\in \Z^d_N} \psi_2 (x) \, \big( e_x - \langle e \rangle_\beta \big) \bigg)
\, + \,
\epsilon^n S \bigg( \sum_{\widetilde{x}\in A_N} \hspace{-0.1cm} \psi_1(\widetilde{x})\, v_{\widetilde{x}} \bigg).
\end{equation*}
Integrating over time and dividing by $t^{1/2}N^{d/2}$, one finds two martingales $M_t$ and $N_t$ such that 
\begin{align*}
\frac{1}{\sqrt t}\int_0^t \mathcal J_N \circ X_s \, \dd s
\; = & \; 
M_t + \frac{1}{t^{1/2} N^{d/2}}  \sum_{\widetilde{x}\in A_N} \hspace{-0.1cm} \psi_1(\widetilde{x})\, \big( u_{\widetilde{x}} - u_{\widetilde{x}} \circ X_t \big) \\
&+ \;
N_t +  \frac{\epsilon^{-1}}{t^{1/2} N^{d/2}}  \sum_{x\in \Z^d_N} \hspace{-0.1cm} \psi_2(x)\, 
\big( (e_{x} - \langle e\rangle_\beta)\circ X_t - (e_{x} - \langle e\rangle_\beta) \big)\\
&+ \; 
\frac{\epsilon^n}{\sqrt t}\int_0^t S  \bigg( \frac{1}{N^{d/2}} \sum_{\widetilde{x}\in A_N} \hspace{-0.1cm} \psi_1(\widetilde{x})\, v_{\widetilde{x}} \circ X_s\bigg) \, \dd s.
\end{align*}
We now separately estimate the variance at equilibrium of the five terms in the right hand side of this equation. 
We remind that, 
by the result of Subsection \ref{subsec: Approximate solution to the Poisson equation}, 
the zero-mean functions $u_{\widetilde{x}}$ and $v_{\widetilde{x}}$ are smooth and local, uniformly in $\widetilde{x}\in A_N$. 
All the estimates below hold almost surely in the limit $N\rightarrow \infty$.

Let us start with the two martingales. First, by \eqref{uniform bound psi 1},
\begin{align*}
\Mean_\beta \big( M_t^2 \big)
\; =& \; 
\bigg\langle 
\bigg(
\frac{1}{N^{d/2}} \sum_{\widetilde{x}\in A_N} \psi_1 (\widetilde{x}) u_{\widetilde{x}}
\bigg)
\bigg(
\frac{-\epsilon^n S}{N^{d/2}} \sum_{\widetilde{x}\in A_N} \psi_1 (\widetilde{x}) u_{\widetilde{x}}
\bigg) \bigg\rangle_\beta \\
\; \le & \;
\epsilon^n
\bigg\langle 
\bigg(
\frac{1}{N^{d/2}} \sum_{\widetilde{x}\in A_N} \psi_1 (\widetilde{x}) u_{\widetilde{x}}
\bigg)^2 
\bigg\rangle_\beta^{1/2}
\bigg\langle 
\bigg(
\frac{-S}{N^{d/2}} \sum_{\widetilde{x}\in A_N} \psi_1 (\widetilde{x}) u_{\widetilde{x}}
\bigg)^2 
\bigg\rangle_\beta^{1/2}
\; = \; \mathcal O (\epsilon^n).
\end{align*}
Next, since $S e_{x} = 0$ for every $x\in\Z_N^d$,
\begin{equation*}
\Mean_\beta \big( N_t^2 \big)
\; = \; 
\bigg\langle
\bigg(\frac{\epsilon^{-1}}{N^{d/2}} \sum_{x\in \Z_N^d} \psi_2 (x) (e_{x} - \langle e \rangle_\beta ) \bigg)
\bigg(\frac{-\epsilon^{n-1}S}{N^{d/2}} \sum_{x\in \Z_N^d} \psi_2 (x) (e_{x} - \langle e \rangle_\beta ) \bigg)
\bigg\rangle_\beta 
\; = \; 0.
\end{equation*}
Let us then consider the two terms not involving the operator $S$.
By \eqref{uniform bound psi 1} and \eqref{uniform bound psi 2} and the stationarity of the measure $\langle \cdot \rangle_\beta$, 
\begin{equation*}
\Mean_\beta \bigg( \frac{1}{t^{1/2} N^{d/2}}  \sum_{\widetilde{x}\in A_N} \hspace{-0.1cm} \psi_1(\widetilde{x})\, \big( u_{\widetilde{x}} - u_{\widetilde{x}} \circ X_t \big) \bigg)^2
\; \le \; 
\frac{2}{t} 
\bigg\langle \bigg(
\frac{1}{N^{d/2}} \sum_{\widetilde{x}\in A_N}
\psi_1(\widetilde{x})\, u_{\widetilde{x}}
\bigg)^2\bigg\rangle_\beta
\; = \; \mathcal O ( t^{-1})
\end{equation*}
and similarly
\begin{equation*}
\Mean_\beta 
\bigg( \frac{\epsilon^{-1}}{t^{1/2} N^{d/2}}  \sum_{x\in \Z_N^d} \hspace{-0.1cm} \psi_2(x)\, \big( e_{x} - e_{x} \circ X_t \big) \bigg)^2
\; = \; 
\mathcal O (\epsilon^{-2} t^{-1}).
\end{equation*}
One finally deals with the last term. 
By a classical bound \cite{kip}, there exists a universal constant $\mathrm C < +\infty$ such that 
\begin{align*}
&\Mean_\beta \bigg( \frac{\epsilon^n}{\sqrt t}\int_0^t S  
\bigg(\frac{1}{N^{d/2}} \sum_{\widetilde{x}\in A_N} \hspace{-0.1cm} \psi_1(\widetilde{x})\, v_{\widetilde{x}} \circ X_s\bigg) \, \dd s \bigg)^2 \\
&\le \; 
\mathrm C \epsilon^{2n}
\bigg\langle
\bigg( \frac{S}{N^{d/2}} \sum_{\widetilde{x}\in A_N} \hspace{-0.1cm} \psi_1(\widetilde{x})\, v_{\widetilde{x}}  \bigg)
(-\epsilon^n S)^{-1}
\bigg( \frac{S}{N^{d/2}} \sum_{\widetilde{x}\in A_N} \hspace{-0.1cm} \psi_1(\widetilde{x})\, v_{\widetilde{x}}  \bigg)
\bigg\rangle_\beta
\; = \; 
\mathcal O (\epsilon^n),
\end{align*}
where \eqref{uniform bound psi 1} has been used to get the last bound. 
The proof is concluded by taking the limit $t\rightarrow \infty$.
$\square$

\subsection{Proving Theorem \ref{the: faster than any power law}}

Since the proof of Theorem \ref{the: faster than any power law} is much similar to that of Theorem \ref{the: faster than any power law bis}, we only indicate the main steps.
We let $n \ge 1$. It is enough to consider the case $m=n$ ; we then need to take the limit of the following quantity: 
\begin{equation*}
\beta^2 \bigg\langle \bigg( \frac{\epsilon}{\sqrt{t}}\int_0^{\epsilon^{-n}t} \mathcal J_N \circ X_s^{(\epsilon,0)} \, \dd s \bigg)^2 \bigg\rangle_\beta.
\end{equation*}
As in the proof of Theorem \ref{the: faster than any power law}, 
we let $\delta > 0$ as fixed in Subsection \ref{subsec: Resonances}, and we let $\alpha, c> 0$ be such that \eqref{measure non resonant set} holds with $r = n$.
We also fix a realization $\omega$ of the frequencies. 
Assuming that $\omega \in D_{\alpha,c}(\widetilde{x},n)$, we write as in Subsection \ref{subsec: Approximate solution to the Poisson equation}, that 
\begin{equation*}
j_{\widetilde{x}} \; = \; - L u_{\widetilde{x}} + \epsilon^n A_{anh} u^{(n)}_{\widetilde{x}} \; =: \; - L u_{\widetilde{x}} + \epsilon^n w_{\widetilde{x}} .
\end{equation*}
All the results of Subsection \ref{subsec: Resonances} still hold, and we combine this with the decomposition \eqref{decomposition current resonances} of the total current, 
to get
\begin{align*}
\frac{\epsilon}{\sqrt t }\int_0^{\epsilon^{-n}t} \mathcal J_N \, \dd s
\; = &\;  
\frac{\epsilon}{\sqrt{t}}\int_0^{\epsilon^{-n}t} \frac{-L}{N^{d/2}} \sum_{\widetilde{x}\in A_N} \psi_1 (\widetilde{x})   u_{\widetilde{x}} \, \dd s
+
\frac{1}{\sqrt{t}}\int_0^{\epsilon^{-n}t} \frac{-L}{N^{d/2}} \sum_{x\in \Z_N^d} \psi_2 (x) \big( e_x - \langle e \rangle_\beta \big) \, \dd s \\
& +
\frac{\epsilon^{n+1}}{\sqrt{t}}\int_0^{\epsilon^{-n}t} \frac{1}{N^{d/2}} \sum_{\widetilde{x}\in A_N} \psi_1 (\widetilde{x}) w_{\widetilde{x}} \, \dd s.
\end{align*}
We then obtain bounds valid almost surely in the limit $N \rightarrow \infty$.
Since the generator $L$ is antisymmetric, the variances of the terms involving the generator $L$ read
\begin{align*}
\bigg\langle \bigg( 
\frac{\epsilon}{\sqrt{t}}\int_0^{\epsilon^{-n}t} \frac{-L}{N^{d/2}} \sum_{\widetilde{x}\in A_N} \psi_1 (\widetilde{x})   u_{\widetilde{x}}
\bigg)^2 \bigg\rangle_\beta
\; & = \; 
\bigg\langle \bigg( 
\frac{-\epsilon}{\sqrt{t}} \frac{1}{N^{d/2}} \sum_{\widetilde{x}\in A_N } \psi_1 (\widetilde{x}) \big( u_{\widetilde{x}} \circ X_{\epsilon^{-n} t} - u_{\widetilde{x}} \big)
\bigg)^2 \bigg\rangle_\beta \\
\; & = \; 
\mathcal O(\epsilon^{2} t^{-1}),\\
\bigg\langle \bigg( 
\frac{1}{\sqrt{t}}\int_0^{\epsilon^{-n}t} \frac{L}{N^{d/2}} \sum_{x\notin A_N} \psi_2 (x) e_x
\bigg)^2 \bigg\rangle_\beta
\; &= \; \mathcal O (t^{-1}).
\end{align*}
Finally, by Jensen's inequality
\begin{align*}
\bigg\langle\bigg( 
\frac{\epsilon^{n+1}}{\sqrt{t}}\int_0^{\epsilon^{-n}t} \frac{1}{N^{d/2}} \sum_{\widetilde{x}\in A_N} \psi_1 (\widetilde{x}) w_{\widetilde{x}} \, \dd s 
\bigg)^2 \bigg\rangle_\beta
\; \le &\; 
\epsilon  t \,
\bigg\langle \bigg( 
\frac{1}{\epsilon^{-n}t} \int_0^{\epsilon^{-n}t} \Big( \frac{1}{N^{d/2}} \sum_{\widetilde{x}\in A_N} \psi_1 (\widetilde{x}) w_{\widetilde{x}} \Big)^2 \, \dd s
\bigg)\bigg\rangle_\beta \\
\; = & \; 
\epsilon  t \,
\bigg\langle \bigg( \frac{1}{N^{d/2}} \sum_{\widetilde{x}\in A_N} \psi_1 (\widetilde{x}) w_{\widetilde{x}} \bigg)^2 \bigg\rangle_\beta
\; = \; 
\mathcal O (\epsilon t) .
\end{align*}
The proof is achieved by performing successively the limits $\epsilon\rightarrow 0$ and then $t\rightarrow \infty$. 
$\square$

\section{Related models}\label{sec: Related models}

We show how to extend our result to two different physical systems, and suggest an analogy with a chain of anharmonic oscillators.
We only deal with the weak coupling limit, and clearly indicate where some pieces of information are missing to get true mathematical statements.

\subsection{Weakly perturbed one-dimensional disordered harmonic chain}
In \cite{dha}, a one-dimensional disordered harmonic chain perturbed by small anharmonic interactions is considered. 
The phase space is the set of points $(q,p)\in \R^{2N}$, and the hamiltonian is of the form
\begin{align}
H (q,p) \; = &\; 
\frac{1}{2}\sum_{x\in\Z_N}  \big( p_x^2 + \omega_x^2 q_x^2 \big) 
+ \frac{1}{2}\sum_{x\in\Z_N} (q_x - q_{x+1})^2
+ \frac{\epsilon}{4} \sum_{x\in \Z_N} q_x^4
+ \frac{\epsilon}{4} \sum_{x\in \Z_N} (q_x - q_{x+1})^4  \label{Dhar Lebowitz chain}\\
\; = &\; H_{har} + \epsilon H_{anh}. \nonumber
\end{align}
The hypotheses on $(\omega_x)_x$ are as in Section \ref{sec: Model}.\footnote{
In \cite{dha}, randomness is on the masses and not on the pinning. This however should not make any crucial difference.
}
The time evolution is predicted by the usual hamiltonian equations.
At $\epsilon = 0$, the chain is harmonic, and any state can be written as a linear combinations of the eigenmodes $(\xi^k)_{1 \le k \le N}$ of the systems ; 
moreover, due to everywhere onsite pinning and since the chain is one-dimensional, all these modes are typically exponentially localized. 
See for example \cite{ber2} and references therein.
Let us also denote by $(\overline{\omega}_k)_{1\le k \le N}$ the eigenfrequencies of the modes, which, as opposed to $(\omega_x)_{x}$,
do not form a sequence of independent and identically distributed variables.
At least in a naive picture of Anderson localization, 
it should be however reasonable to think that the frequency $\overline{\omega}_1$ of a mode $\xi_1$ localized near a point $x_1\in\Z_N$, 
and a frequency $\overline{\omega}_2$ of a mode $\xi_2$ localized near a point $x_2\in\Z_N$, 
quickly decorrelate as $|x_1-x_2|$ grows. 
This should be needed to get rid of resonances as we did in Subsection \ref{subsec: Resonances}.

In this model as well, the generators $A_{har}$ and $A_{anh}$ exchange $p$-symmetric and $p$-antisymmetric functions. 
Now, the crux of the matter is that, for a function $f$ that is $p$-antisymmetric, and that only depends on variables around some site $x$, the Poisson equation 
\begin{equation*}
- A_{har} u \; = \; f
\end{equation*}
should be solved in the same way as we did, with a solution $u$ that is exponentially localized near $x$. 
Let us mention that this equation has been solved in \cite{ber2} in the case where $f$ is the current corresponding to a harmonic interaction. 
This case is quite special however, since there, a lucky cancellation makes disappear all possible resonances. 

Looking at the proof of Lemma \ref{lem: Poisson equation harmonic}, one has indeed 
\begin{equation*}
u(q,p) \; = \; \lim_{0 < z \rightarrow 0} \int_0^{\infty} \ed^{-zt} f \circ X_t^{har} (q,p) \, \dd t. 
\end{equation*}
This can be made more explicit since, using (4.4) and (4.5) in \cite{ber2}, one can write 
\begin{align*}
 f \circ X_t^{har} (q,p)
\; =& \; 
f \bigg( \dots, 
\sum_{j=1}^N \Big( \langle q , \xi^j \rangle \cos \overline{\omega}_j t  + (\langle p , \xi^j \rangle / \overline{\omega}_j ) \sin \overline{\omega}_j t \Big), \\
&\; \phantom{f \bigg( \dots,}
\sum_{j=1}^N \Big( - \overline{\omega}_j \langle q , \xi^j \rangle \cos \overline{\omega}_j t  + \langle p , \xi^j \rangle  \sin \overline{\omega}_j t \Big),
\dots \bigg).
\end{align*}
Here as well, one can define $g(q,p; \theta)$ by replacing $\overline{\omega} t$ by $\theta$ in this last formula. 
Defining then an angle $\overline{\theta}$ such that $\overline{\theta}_j$ is the phase of $\langle q , \xi^j \rangle + i \langle p , \xi^j \rangle / \overline{\omega}_j$, 
one then finds that the $p$-antisymmetry of $f$ translates into the relation $g (q,p, 2\overline{\theta} - \theta) = - g (q,p, \theta)$, 
which ensures that the Fourier expansion has no constant mode. 
Finally, the exponential localization of the eigenmodes should guarantee that only a small number of eigenfrequencies are practically involved, 
so that diophantine estimates can be used.
Moreover, while the solution $u$ itself will not be local anymore, it will be exponentially localized,
and the limit $\epsilon \rightarrow 0$ should allow us to get rid of exponential tails. 

\Remark
For the chain defined by \eqref{Dhar Lebowitz chain}, 
it is known \cite{aok} that the Green-Kubo conductivity $\kappa (\beta,\epsilon)$, 
seen as function of the inverse temperature $\beta$ and the coupling strength $\epsilon$, satisfies the exact scaling
\begin{equation*}
\kappa (\beta, \epsilon) \; = \; \kappa (\beta/r,\epsilon /r) \quad \text{for any} \quad r >0.
\end{equation*}
Therefore, the behavior of the conductivity in the small coupling regime corresponds to its behavior at low temperatures.
Many numerical works deal with the diffusion of an initially localized energy packet (diffusion at ``zero temperature"). 
The heat equation can sometimes serve as a good phenomenological model to describe the spreading of this packet \cite{mul}. 
If one assumes that the diffusion constant in this equation is given by the Green-Kubo conductivity\footnote{
While this looks reasonable, it is fair to say that this identification is not justified, since we do not look at the system in a true macroscopic scale.}, 
then our results predict in fact a subdiffusive spreading of energy slower than any power law.

\subsection{Disordered classical spin chain}
In \cite{oga}, a one-dimensional chain of classical spins is studied. 
Let $\mathbb S$ be the unit sphere in $\R^3$, 
and let a point of the phase space be given by $\mathbf S = (\mathbf S_x)_{x\in\Z_N} \in \mathbb S^N$.
The hamiltonian $H$ is given by 
\begin{equation*}
H (\mathbf S) \; = \; \sum_{x\in\Z_N} \boldomega_x \cdot \mathbf S_x + \epsilon \sum_{x\in\Z_N} \mathbf S_x \cdot \mathbf S_{x+1}, 
\end{equation*}
and the equations of motions read
\begin{equation}\label{equation of motion Oganesyan Pal Huse}
\dot{\mathbf S}_x \; = \; \nabla_x H \wedge \mathbf S_x, 
\end{equation}
where $\nabla_x = (\partial/ \partial S_{1,x}, \partial/ \partial S_{2,x} ,\partial/ \partial S_{3,x})$, 
where $\mathbf S \cdot \mathbf T$ is the standard scalar product in $\R^3$, 
and where $\mathbf S \wedge \mathbf T$ is the standard vector product in $\R^3$.
It is assumed that $(\boldomega_x)_{x\in \Z_N}$ form a sequence of independent and identically distributed random vectors in $\R^3$.

The dynamics \eqref{equation of motion Oganesyan Pal Huse} is generated by $A$, defined as
\begin{align*}
A u 
\; & = \; 
\sum_{x\in \Z_N} (\nabla_x H \wedge \mathbf S_x ) \cdot \nabla_x u
\; = \; 
\sum_{x\in \Z_N} (\boldomega_x \wedge \mathbf S_x ) \cdot \nabla_x u
+ \epsilon \sum_{x\in\Z_N} \big( (\mathbf S_{x-1} + \mathbf S_{x+1}) \wedge \mathbf S_x \big) \cdot \nabla_x u \\
\; & = \; 
A_{har} u + \epsilon A_{anh} u
\end{align*}
where the denominations $A_{har}$ and $A_{anh}$ are aimed to bare the analogy with our model. 
The energy of the particle at site $x$ is defined as 
\begin{equation*}
e_x \; = \; \boldomega_x \cdot \mathbf S_x + \frac{\epsilon}{2} \mathbf S_{x-1}\cdot \mathbf S_x + \frac{\epsilon}{2} \mathbf S_x \cdot \mathbf S_{x+1},
\end{equation*}
and the relation 
\begin{equation*}
A e_x \; = \; \epsilon \big( j_{x-1,x} - j_{x,x+1} \big)
\end{equation*}
is satisfied if one defines the current $j_{x,x+1}$ by 
\begin{equation*}
j_{x,x+1} \; = \; 
\frac{\boldomega_x + \boldomega_{x+1}}{2}\cdot (\mathbf S_x \wedge \mathbf S_{x+1})
+ \frac{\epsilon}{2} \big( (\mathbf S_{x-1}\wedge \mathbf S_x)\cdot \mathbf S_{x+1} + (\mathbf S_x\wedge \mathbf S_{x+1})\cdot \mathbf S_{x+2} \big).
\end{equation*}

At $\epsilon = 0$, each spin just precesses around the axis characterized by the vector $\boldomega_x$, at a constant angular velocity $|\boldomega_x|_2$. 
Its energy $\boldomega_x \cdot \mathbf S_x$ is conserved, and determines the plane perpendicular to $\boldomega_x$ where the precession takes place. 
The picture in phase space is thus here completely analogous to that of the one-dimensional harmonic oscillators we have considered. 

To proceed as before, we would now need to partition the set of functions on the phase space into two subspaces, 
playing the role of $p$-symmetric and $p$-antisymmetric functions. 
This is easily made possible if we add the extra assumption that, writing $\boldomega_x = (\omega_{1,x},\omega_{2,x},\omega_{3,x})$, one has almost surely
\begin{equation}\label{extra assumption magnetic field}
\omega_{1,x} \; = \; 0 \qquad \text{for all } x \in \Z_N .
\end{equation}
We then define $1$-symmetric functions on $\mathbb S^N$ as functions that are 
symmetric under the reflection that simultaneously maps $\omega_{1,x}$ to $-\omega_{1,x}$  for every $x\in\Z_N$ ; 
$1$-antisymmetric functions are defined similarly. 
Then the currents $j_{x,x+1}$ are $1$-antisymmetric functions, and the generator $A$ maps $1$-symmetric functions to $1$-antisymmetric functions and vice versa. 
This is all what is needed to compute a finite number of perturbative steps as we did. 
Without the assumption \eqref{extra assumption magnetic field}, 
we have not been able to find a symmetry that guarantees the solvability of the hierarchy \eqref{perturbation order 1}-\eqref{perturbation order 2}.
Still, such a symmetry is noway a necessary condition ; we believe that this is only a technical question, and that these equations can in fact be solved.

\subsection{One dimensional chain of strongly anharmonic oscillators}

We finally suggest an analogy with a a one-dimensional chain of identical strongly anharmonic oscillators. 
The phase space is the set of points $(q,p)\in \R^{2N}$, and the hamiltonian is given by 
\begin{equation*}
H(q,p) \; = \; \sum_{x\in\Z_N} \Big( \frac{p^2_x}{2}+ \frac{q_x^4}{4} \Big) + \frac{\epsilon}{2} \sum_{x\in\Z_N} (q_x - q_{x+1})^2.
\end{equation*}
Equations of motion are the usual Hamilton equations. 
For such a system, existence of breathers is well known in the infinite volume limit \cite{mac}.
Unfortunately, this as such does not say us much about the conductivity of the chain. 

Let us see that, at least in the asymptotic regime we consider, the effect of the strongly anharmonic onsite potentials could be compared to that of random frequencies.
At $\epsilon = 0$, all the oscillators are characterized by an energy $e_{x,0}$, and evolve independently from each other. 
Since their dynamics is one-dimensional, they are just oscillating periodically at a frequency $\omega_{x,0} \sim ( e_{x,0} )^{1/4}$. 
So, as far as the uncoupled dynamics is concerned, 
a typical state of a chain of identical anharmonic oscillators on the one hand, 
and a typical state of a disordered chain of harmonic oscillators on the other hand, are qualitatively completely similar.
In the first case however, typical means typical with respect to the Gibbs measure, 
while in the second it means typical with respect to both the Gibbs measure and the disorder.

However, when trying to adapt our strategy, the problem of resonances cannot be ruled out so easily.
Indeed, in the disordered chain, resonances only affected some fixed sites, while here, the places where resonances occur also move with time, 
a situation which in fact favors the transport of energy. 
Still our main intuition remains that, 
due to quick averaging of fast oscillations, energy gets trapped in finite portions of space for time-scales that might not be an inverse polynomial of $\epsilon$.

\vspace{0.5cm}

\noindent
\textbf{Acknowledgements.} 
It is a pleasure for me to warmly thank C\'edric Bernardin and Wojciech De Roeck for encouragements and stimulating discussions.
I thank the European Advanced Grant Macroscopic Laws and Dynamical Systems (MALADY) (ERC AdG 246953) for financial support.


\end{document}